\documentclass[%
reprint,
superscriptaddress,
amsmath,amssymb,
aps,
]{revtex4-2}

\usepackage{graphicx}
\usepackage{dcolumn}
\usepackage{bm}
\usepackage[separate-uncertainty = true,multi-part-units=single]{siunitx}
\usepackage{xspace}
\usepackage[version=4]{mhchem}
\usepackage[colorlinks,
linkcolor=blue,
anchorcolor=blue, 
citecolor=blue,
urlcolor=blue,
]{hyperref}

\newcommand{\tplus}{\ensuremath{^{3+}}\xspace}
\newcommand{\erg}{\ensuremath{^4}I\ensuremath{_{15/2}}\xspace}
\newcommand{\ere}{\ensuremath{^4}I\ensuremath{_{13/2}}\xspace}
\begin{document}
\preprint{APS/123-QED}

\title{Spectral broadening of a single Er$\ensuremath{^{3+}}\xspace$ ion in a Si nano-transistor}
\author{Jiliang Yang}
\author{Jian Wang}
\author{Wenda Fan}
\author{Yangbo Zhang}
\author{Changkui Duan}
\affiliation{CAS Key Laboratory of Microscale Magnetic Resonance and School of Physical Sciences, University of Science and Technology of China, Hefei 230026, China}
\affiliation{CAS Center for Excellence in Quantum Information and Quantum Physics, University of Science and Technology of China, Hefei 230026, China}
\author{Guangchong Hu}
\author{Gabriele G. de Boo}
\affiliation{Centre of Excellence for Quantum Computation and Communication Technology, School of Physics, University of New South Wales, NSW 2052, Australia}
\author{Brett C. Johnson}
\affiliation{Centre of Excellence for Quantum Computation and Communication Technology, School of Engineering, RMIT University, Victoria 3001, Australia}
\affiliation{Centre of Excellence for Quantum Computation and Communication Technology, School of Physics, University of Melbourne, Victoria 3010, Australia}
\author{Jeffrey C. McCallum}
\affiliation{Centre of Excellence for Quantum Computation and Communication Technology, School of Physics, University of Melbourne, Victoria 3010, Australia}
\author{Sven Rogge}
\affiliation{Centre of Excellence for Quantum Computation and Communication Technology, School of Physics, University of New South Wales, NSW 2052, Australia}
\author{Chunming Yin}
\email{Chunming@ustc.edu.cn}
\author{Jiangfeng Du}
\affiliation{CAS Key Laboratory of Microscale Magnetic Resonance and School of Physical Sciences, University of Science and Technology of China, Hefei 230026, China}
\affiliation{CAS Center for Excellence in Quantum Information and Quantum Physics, University of Science and Technology of China, Hefei 230026, China}
\date{\today}
\begin{abstract}
Single rare-earth ions in solids show great potential for quantum applications, including single photon emission, quantum computing, and high-precision sensing. However, homogeneous linewidths observed for single rare-earth ions are orders of magnitude larger than the sub-kilohertz linewidths observed for ensembles in bulk crystals. The spectral broadening creates a significant challenge for achieving entanglement generation and qubit operation with single rare-earth ions, so it is critical to investigate the broadening mechanisms. We report a spectral broadening study on a single Er\tplus ion in a Si nano-transistor. The Er-induced photoionisation rate is found to be an appropriate quantity to represent the optical transition probability for spectroscopic studies, and the single ion spectra display a Lorentzian lineshape at all optical powers in use. Spectral broadening is observed at relatively high optical powers and is caused by spectral diffusion on a fast time scale.
\end{abstract}
\maketitle
\section{Introduction}
Rare-earth ions in solids are promising candidates for quantum repeater and quantum memory applications due to their extraordinarily long spin coherence times~\cite{zhong_optically_2015,Coherence2018} and narrow homogeneous linewidths. Sub-kilohertz homogeneous linewidths of the 4f$-$4f transitions have been observed for rare-earth ensembles in bulk crystals via photon echo measurements~\cite{bottger_effects_2009, Subkilohertz2020}, limited by dephasing processes due to magnetic or electric field fluctuations in the local environment, spin flip-flops, and spin-phonon relaxation~\cite{Opticaldecoherence,Dephasing_ceramics_16,Subkilohertz2020}. However, interactions between rare-earth ions and their dynamic environment cause the emission line of each ion to diffuse around an equilibrium spectral position randomly. These spectral diffusion processes would impact the phase coherence for both ensemble and single rare-earth ion applications~\cite{Asadi_2020,Spectral_Er_21}, so it is important to study and control these processes for practical applications based on rare-earth ions. For crystalline materials at low temperatures, the dominant interactions responsible for spectral diffusion include Stark shift fluctuations, which can arise from charge fluctuations in proximity to the ions~\cite{Effect_qd_00,Spectral_00} or on the surface of nanoparticles~\cite{bartholomew_optical_2017}, and spin fluctuations via magnetic dipole-dipole interaction~\cite{bottger_optical_2006}. These interactions result in a time-dependent effective homogeneous linewidth, and have been investigated for rare-earth ensembles in both bulk crystals~\cite{bottger_optical_2006, Opticaldecoherence,veissier_optical_2016,Subkilohertz2020} and nanoparticles~\cite{bartholomew_optical_2017}.

Recent progress on detection of single rare-earth ions promotes their potential applications in single photon emission~\cite{zhong_18_Nd,Dibos-PRL-2018, Spectral_Er_21}, quantum computing~\cite{raha_optical_2020,Kindem2020}, and high-precision sensing~\cite{ZhangQ2019}. However, homogeneous linewidths observed for single rare-earth ions are on the order of 1-\SI{10}{\MHz}~\cite{Yinnat2013,utikal_spectroscopic_2014,Dibos-PRL-2018,zhong_18_Nd, Kindem2020,Xia_2020}, considerably larger than the sub-kilohertz homogeneous linewidths observed for ensembles in bulk crystals. This broadening creates a significant challenge for entanglement generation and qubit operation with single rare-earth ions. The narrow 4f$-$4f transitions observed for ensembles correspond to excited states with a long lifetime, resulting in weak fluorescence. As a result, the single ion detection approaches rely on signal enhancement methods such as single-electron charge sensing~\cite{Yinnat2013}, utilisation of higher lying states~\cite{utikal_spectroscopic_2014, Xia_2020}, and Purcell enhancement~\cite{Dibos-PRL-2018,zhong_18_Nd, Kindem2020,Spectral_Er_21}. These inevitably introduce surface and interface states and/or strong optical fields which would perturb the single ions or their environment, in contrast to the well-controlled bulk environment and pump/probe conditions in the ensemble experiments. Identifying the dominant broadening source is key to reduce the linewidth significantly but is yet to be achieved for each detection approach. For example, work on an individual Yb$^{3+}$ ion in a YVO nano-cavity revealed spectral diffusion over \SI{1}{\mega\hertz} on a minute time scale and a slightly smaller optical dephasing limited linewidth~\cite{Kindem2020} while the source of the optical dephasing is not determined.

\begin{figure*}
	\centering\includegraphics[width=1\textwidth]{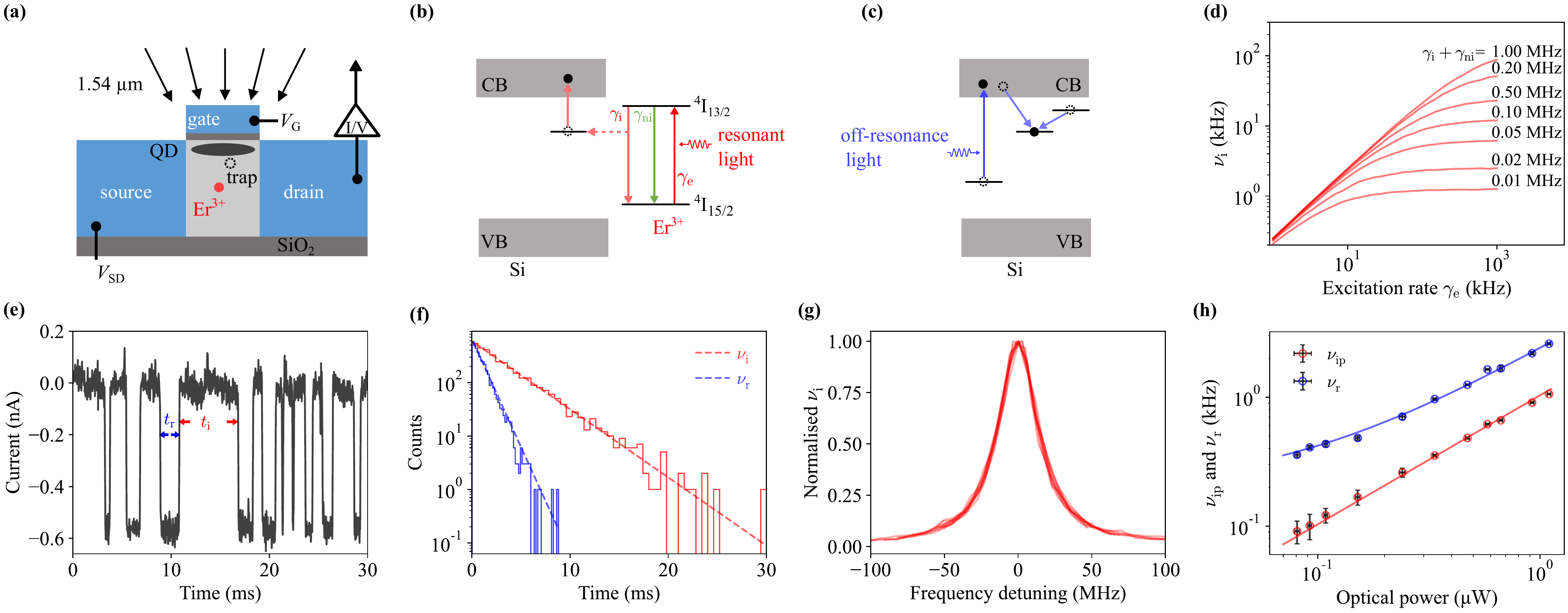}
	\caption{Single Er\tplus ion spectroscopy and continuous-wave (CW) measurement. (a) Cross-section schematic of the FinFET with device connection. The laser light is focused on the Er-doped channel region of the device. Energy level diagrams showing (b) photoionisation and (c) reset processes. CB and VB stand for conduction band and valence band respectively. (d) Simulated relationship between the excitation rate, $\gamma_\textrm{e}$, and the photoionisation rate, $\nu_\mathrm{i}$, under different total decay rates from \ere to \erg. (e) A short section of a typical current$-$time trace under resonant CW laser illumination. $t_\textrm{i}$ and $t_\textrm{r}$ represent the time intervals that it takes to ionise and reset the single trap, respectively. (f) Statistics of $t_\mathrm{r}$ (blue) and $t_\mathrm{i}$ (red) time intervals from a long time trace similar to (e). Both histograms are fitted with an exponential decay function and return a ionisation rate, $\nu_\mathrm{i}$, and a reset rate, $\nu_\mathrm{r}$. (g) Normalised single ion spectra measured under CW optical powers ranging from 0.08 to 1.1~\si{\micro\watt}. (h) Power dependence of the peak ionisation rate $\nu_\mathrm{ip}$ (red circle) and the reset rate $\nu_\mathrm{r}$ (blue circle). Both are well fitted with a linear function (solid lines). The blue line curves in the low power region of the log-log plot due to a zero-power offset that corresponds to a spontaneous reset rate when no laser illumination is present.}
	\label{fig:fig1}
\end{figure*}

This work focuses on the single rare-earth ion detection approach using a hybrid optical-electrical method enabled by single-electron charge sensing~\cite{Yinnat2013}. This approach combines the \SI{1.54}{\micro\meter} telecom-band optical transition of Er\tplus ions with well-established nano-fabrication technologies for Si-based materials. Previously observed linewidths using this approach are on the order of \SI{10}{\MHz}~\cite{Yinnat2013, ZhangQ2019}, and the spectra generally display an asymmetrical lineshape attributed to a combined effect due to charge accumulation and the resulting Stark shift~\cite{Yinnat2013}. As a result, the average electric current, from which the Er spectrum is formed, is related to the optical transition probability for the Er ion under the excitation and electrical measurement conditions used but does not represent the intrinsic excitation rate and spectral lineshape. A more recent study introduces a time-resolved single ion detection method and opens up the possibility for a wide range of excitation conditions to be explored\cite{hu_time-resolved_2022}.

In the work presented here, the time-resolved method is employed to study the spectral broadening of a single Er\tplus ion in a Si nano-transistor. The photoionisation rate of a single trap induced by Er relaxation is found to be an appropriate quantity to study the spectral properties, based on both simulation and experimental results from the optical power dependence. The observed Er spectra show a Lorentzian shape. Spectral broadening is observed at relatively high optical power, and is attributed to spectral diffusion on a time scale shorter than a few microseconds. The photoionisation rate shows a linear dependence on the effective optical power, consistent with a single-photon Er excitation process.

\section{Methods} 
The device used in this study was a Si fin-field-effect transistor (FinFET) consisting of three electrodes and a nanowire channel (\SI{35}{\nano\metre} width $\times$ \SI{80}{\nano\metre} length $\times$ \SI{60}{\nano\metre} height), as illustrated in Fig.~\hyperref[fig:fig1]{1(a)}. A small number ($\sim25$) of $^{167}$Er atoms (nuclear spin, $I = 7/2$) and a larger number ($\sim150$) of O atoms were implanted into the device, and then the device was thermally processed to recover the implantation damage and activate the Er\tplus ions~\cite{Yinnat2013}. The experiment was carried out in a closed-cycle cryostat at \SI{3.6}{\kelvin}. The FinFET was biased under a sub-threshold condition~\cite{SellierSubthreshold} so that quantum dots (QDs) could form in the device channel. Current arising from the electron tunnelling through QDs was amplified by an I-V converter with a \SI{10}{\kHz} analog bandwidth located at room temperature. The measurement conditions allowed individual Er ions to be addressed as per our earlier published work~\cite{Yinnat2013}.

A cavity locked laser was used to excite a single Er\tplus ion with short-term frequency stability below \SI{2}{\mega\hertz}. The long-term cavity drift was monitored with a wavemeter (Bristol 621A-NIR) to calibrate the laser frequency. The specified repeatability of the wavemeter is \SI{+-6}{\MHz}. An electro-optic modulator (EOM) was used to generate two frequency-tunable sidebands while the carrier band was suppressed to less than 0.5$\%$.
By choosing a suitable cavity resonance frequency the laser was locked to, one sideband was used to scan a well separated Er peak for the spectral broadening study. The other sideband and the carrier band were tuned away from any Er resonances. As a result, the laser light consisted of a resonant component for the spectral scan and off-resonance components. The proportion ($\alpha$) of the resonant power to the total optical power had a maximum value of slightly below 0.5 and could be reduced by adding off-resonance light from another laser.

Then the laser light passed through an acousto-optic modulator (AOM) for pulse generation with an extinction ratio of 50~dB and was split into two beams by a fibre optic coupler. One beam passed through a single mode optical fibre and was focused on the device surface with a spot size of $\sim$\SI{3}{\micro\meter} using fibre-based confocal microscopy~\cite{Fiberconfocal}. The other beam was sent to a photodetector for power monitoring. Unless expressly stated, the optical power used in this work refers to the total optical power before the fibre port on top of the cryostat with $\alpha =$ \num{0.496(2)}. Fibre polarisation controllers and a fibre polariser were used to tune the light polarisation for the EOM and to set up a polarisation state for the device. Slow polarisation drift due to the two fibre polarisation controllers before the fibre polariser led to a slight optical power fluctuation. This power fluctuation was monitored by the photodetector for each measurement and is presented as a standard deviation in the following figures.

\section{Results and Discussions}
Figures~\hyperref[fig:fig1]{1(b), (c)} show the processes involved in the single ion spectroscopy. The Er$^{3+}$ ion is first resonantly excited from the ground state (\erg) to the excited state (\ere), and it relaxes back to the ground state via either a radiative process by emitting a photon, or non-radiative processes such as Auger effect~\cite{Erbium_in_silicon05}. The charge ionisation due to an Auger process affects the chemical potential of the QD, which in turn changes the tunnelling current through the QD. In this device, a single trap in close proximity to the QD is ionised by Auger relaxation of the Er\tplus ion and induces a significant change in the tunnelling current. Then the ionised trap will be reoccupied (reset) by capturing an electron from a nearby reservoir or from photon-generated carriers in the channel, and consequently, the tunnelling current resets to the original level.

Among the processes shown in Fig.~\hyperref[fig:fig1]{1(b), (c)}, the excitation process, at a rate of $\gamma_\textrm{e}$, is a straightforward choice for spectral broadening and linewidth studies but is not directly observable. However, the photoionisation process can be observed by monitoring the change in the current and is used here for spectral study under appropriate conditions. To investigate the relationship between $\gamma_\textrm{e}$, and the photoionisation rate, $\nu_\mathrm{i}$, Monte-Carlo simulations are carried out with different presumed total decay rates. The total decay rate is the sum of the single-trap ionisation decay rate, $\gamma_\textrm{i}$, and the combined rate of all the other decay processes, $\gamma_\textrm{ni}$. The simulation results in both Fig.~\hyperref[fig:fig1]{1(d)} and Fig.~S2 in the Supplemental Material~\cite{supp_mat} suggest that $\nu_\mathrm{i}$ is approximately in proportion to $\gamma_\textrm{e}$ as long as $\gamma_\textrm{e}$ is well below the total decay rate. The total decay rate in this device is determined to be higher than \SI{1}{\MHz} (see Section I in the Supplemental Material~\cite{supp_mat}), and thus is much higher than the expected excitation rate under the relatively low light intensity used in this study. Therefore, the ionisation rate, $\nu_\mathrm{i}$, is used to represent the excitation rate in the following study. Besides, the lower bound on the total decay rate also puts a lower bound on the linewidth.

The spectral broadening study in this work is carried out in two modes, continuous-wave (CW) laser excitation and pulsed laser excitation. A common side effect observed in the FinFETs is that laser illumination introduces current noise and heating, which lowers the electrical sensitivity of the current signal. In order to avoid this, CW laser excitation is used in a low optical power regime, and pulsed laser excitation is used at high optical powers. These settings ensure the observation of a large number of photoionisation events within a reasonable measurement time at each laser frequency.

The spectral broadening is first investigated under CW excitation. The maximum CW power is limited to \SI{1.1}{\micro\watt} to maintain a sufficient signal contrast. The laser frequency is scanned over a spectral peak for each optical power, and a \SIrange{20}{60}{\s} current-time trace is recorded at each frequency.  Figure~\hyperref[fig:fig1]{1(e)} shows a short section of a typical time trace of current when the device is under CW resonant excitation with an optical power of \SI{0.33}{\micro\watt}. The current switches between two discrete levels, and the high level and low level correspond to the occupied trap state and the ionised trap state, respectively. For each ionisation-reset event, an ionisation time $t_\textrm{i}$ and a reset time $t_\textrm{r}$ are determined as shown in Fig.~\hyperref[fig:fig1]{1(e)}. The statistics of a large number of events from a long current-time trace is plotted as the two histograms in Fig.~\hyperref[fig:fig1]{1(f)}, and both follow an exponential decay function. The fittings give an ionisation rate of $\nu_\mathrm{i} =$ \SI{294(5)}{\Hz}, and a reset rate of $\nu_\mathrm{r} =$ \SI{929(14)}{\Hz}. These two rates are measured as the laser scans the Er resonant peak under different optical powers, and the ionisation rate, $\nu_\textrm{i}$, as a function of laser frequency detuning is plotted as one spectral line for each optical power in Fig.~\hyperref[fig:fig1]{1(g)}. For comparison purposes, each spectral line is normalised against its highest point and detuned in frequency against its peak position.  All the spectra can be well fitted with a Lorentzian lineshape, and the fittings give a near-constant linewidth of \SI{32(2)}{\MHz} (full width at half maximum). The spectra with absolute ionisation rates can be found in Section IV of the Supplemental Material~\cite{supp_mat}. The fitted peak position fluctuates up to \SI{15}{\MHz} in this work without any clear dependence on the experimental conditions. This fluctuation is close to the frequency repeatability of the wavemeter and thus is not investigated further in this work.

The reset rate (blue circle), $\nu_\mathrm{r}$, and the fitted peak height, defined as the peak ionisation rate (red circle), $\nu_\mathrm{ip}$, are plotted as a function of optical power in Fig.~\hyperref[fig:fig1]{1(h)}. The peak ionisation rate is in proportion to the optical power with a slope of \SI[per-mode = symbol]{1.03(1)}{\kHz\per\uW}. This proportional relationship indicates a single-photon Er excitation process and verifies the simulated results in Fig.~\hyperref[fig:fig1]{1(d)}. The reset rate also shows a linear dependence on the optical power, and a linear fit (blue line) gives a slope of \SI[per-mode = symbol]{2.21(5)}{\kHz\per\uW}. The zero-power offset corresponds to a spontaneous reset rate of \SI[per-mode = symbol]{0.199(29)}{\kHz} when no laser illumination is present~\cite{hu_time-resolved_2022}. 
\begin{figure}
	\centering\includegraphics[width=0.45\textwidth]{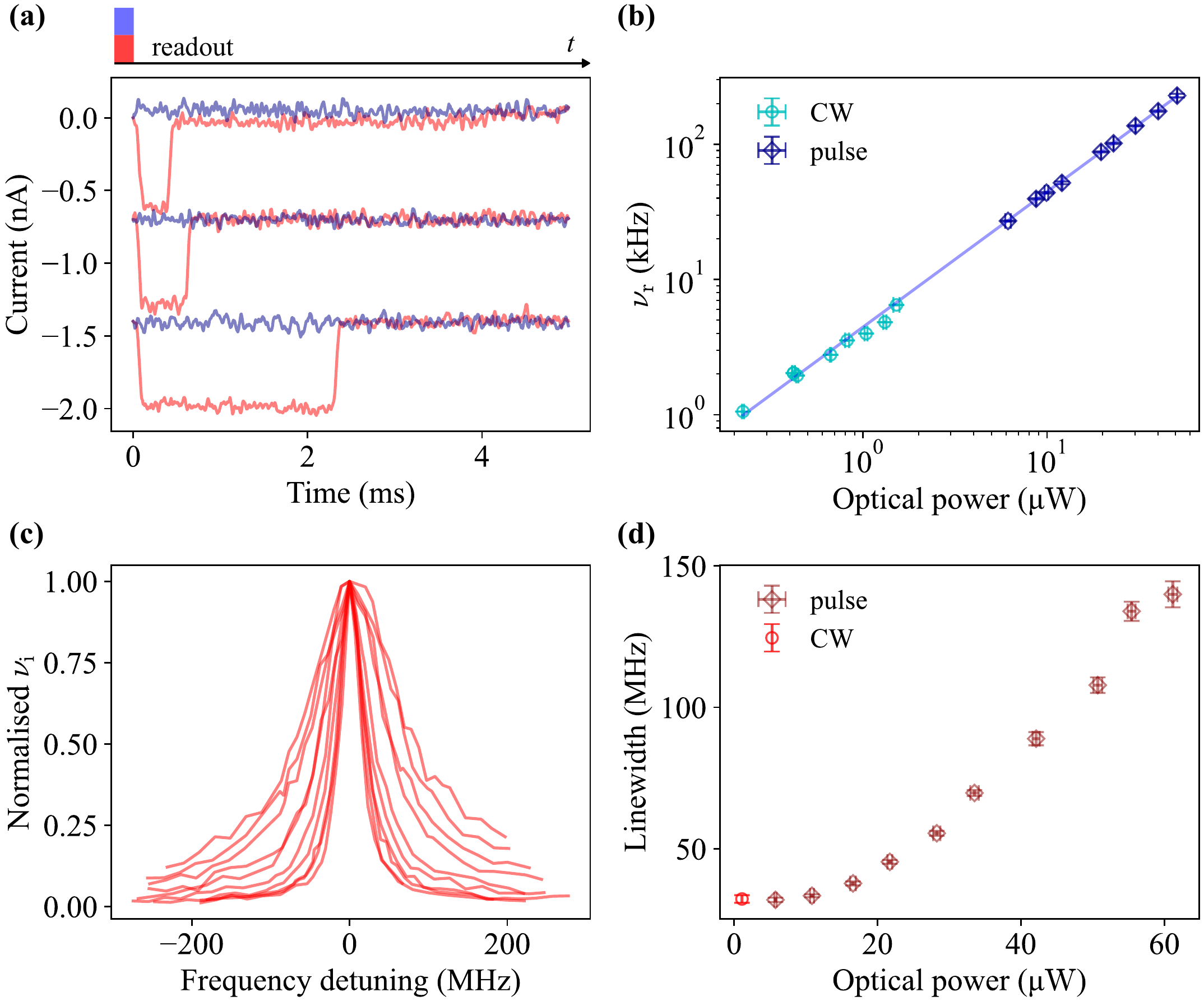}
	\caption{\label{fig:fig2}Pulsed single Er\tplus ion spectroscopy. (a) Pulse sequence and typical current-time traces. The red and blue rectangles in the laser pulse represent the resonant and off-resonance components, respectively. The three pairs of traces are offset along y-axis for clarity. The red traces with current falling below a threshold level in the readout window are identified as successful ionisation counts, while the blue traces staying above the threshold are counted as idle cycles. (b) Power dependence of the reset rate. The data from both CW (cyan circle) and pulsed (blue diamond) measurements follows a linear relationship (solid line) as the optical power varies by two orders of magnitude. (c) Normalised spectra from pulsed measurements under different optical powers ranging from \SIrange{5.8}{61}{\uW}. (d) Power dependence of the spectral linewidth from both CW (red circle) and pulsed (brown diamond) measurements.}
\end{figure}

The spectral broadening at higher optical power is investigated with pulsed laser excitation. A short pulse length is chosen to ensure a negligible impact of the laser pulse on the electrical sensitivity, and there is a long time delay after the pulse for an ionisation count to reset. Specifically, each measurement cycle consists of a \SI{4}{\micro\second} excitation pulse and a \SI{5}{\milli\second} dark period, as is shown in the top panel of Fig.~\hyperref[fig:fig1]{2(a)}. The measurement cycle is repeated 20,000 - 50,000 times at each laser frequency. Then the photoionisation events in all valid cycles are determined by comparing the current signal in the readout window to a threshold. Several cycles are shown in Fig.~\hyperref[fig:fig2]{2(a)}, and the current immediately before the laser pulse is subtracted from each trace as an offset. The red traces are identified as photoionisation counts as the current falls below the threshold in the readout window. In contrast, the blue traces are idle cycles as the current signal remains above the threshold. Furthermore, a small proportion of traces show a below-threshold signal before the laser pulse due to an unsuccessful reset from the previous cycle, and these traces are counted as invalid cycles. Finally, photoionisation probability, defined as the ratio of photoionisation counts to the total valid cycles, is calculated for each group of cycles measured under the same condition.

As shown in the previous study~\cite{hu_time-resolved_2022}, the ionisation and reset processes can be described by a two-state Markov model. The photoionisation probability, $R$, can be expressed as:
\begin{equation} 
	\label{eqn:eq1}
	R =\frac{\nu_\mathrm{i}}{\nu_\mathrm{i}+\nu_\mathrm{r}}(1-e^{-(\nu_\mathrm{i}+\nu_\mathrm{r})t_\mathrm{p}}),
\end{equation}
where $t_\mathrm{p}$ is the pulse length, and $\nu_\mathrm{r}$ and $\nu_\mathrm{i}$ are the reset rate and the ionisation rate respectively.

In order to calculate the ionisation rate from the measured photoionisation probability, the reset rate needs to be determined first. The reset rate under a high optical power can be measured with a two-pulse measurement (see Section III in the Supplemental Material~\cite{supp_mat}). The results from CW and pulsed measurements with the same experimental setup show a linear dependence of the reset rate on the optical power (Fig.~\hyperref[fig:fig2]{2(b)}). However, adding a second laser that is required for the two-pulse measurement would change the polarisation, which in turn affects the actual light intensity in the channel of the device and the reset rate. To avoid setup modification, measurements with two lasers are performed after single-laser measurements. For polarisation consistency, reset rates used for calculating the ionisation rate are extrapolated from the linear dependence shown in Fig.~\hyperref[fig:fig1]{1(h)}.

Next, the ionisation rate is calculated for each measured photoionisation probability based on Eq.~\hyperref[eqn:eq1]{(1)} and the extrapolated reset rate. Figure~\hyperref[fig:fig2]{2(c)} show the spectra measured under different optical powers ranging from \SIrange{5.8}{61}{\uW}. For comparison purposes, the ionisation rates are normalised against the maximum rate at each power, and the same spectra with absolute ionisation rates can be seen in Section IV in the Supplemental Material~\cite{supp_mat}). All lines are well fitted with a Lorentzian lineshape, and the linewidths are plotted as a function of optical power in Fig.~\hyperref[fig:fig2]{2(d)}. At low optical powers, the linewidths measured with pulsed laser excitation are close to the linewidth from the CW measurement (red circle) but start to increase as the optical power increases.
\par
The observed power broadening may arise from photo-induced processes that do not require the excitation of the Er\tplus ion. This broadening would grow with the total optical power. On the other hand, instantaneous spectral diffusion~\cite{huang_excess_1989, liu_laser-induced_1990} may also lead to power broadening with a strong dependence on the resonant power. To further investigate the underlying mechanism, two additional experiments are carried out.

\begin{figure}[t]
	\centering\includegraphics[width=0.45\textwidth]{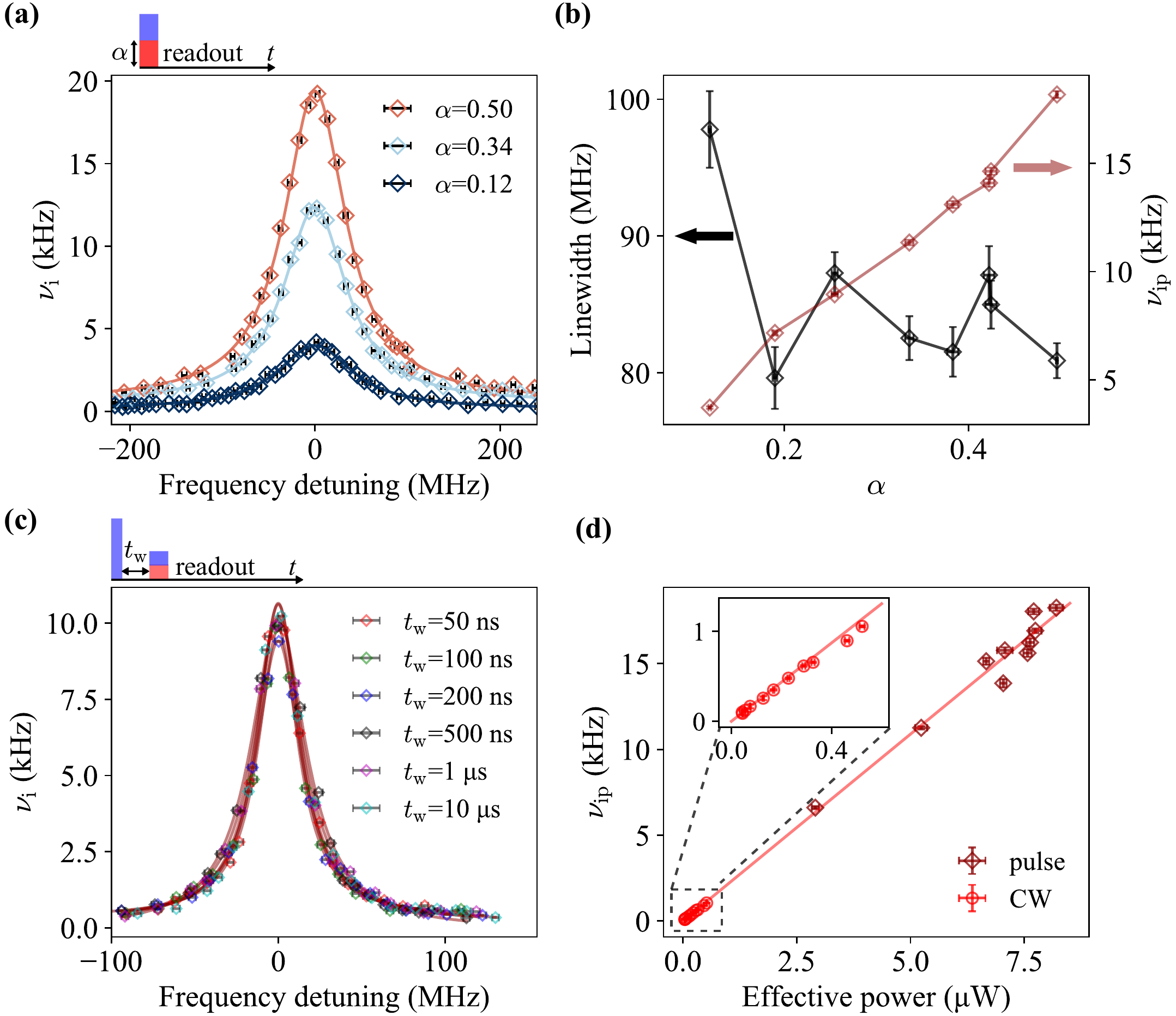}
	\caption{Power broadening mechanism study. (a) Pulse sequence for the impact of resonant power and three example spectra. Here $\alpha$ represents the proportion of the resonant power to the total optical power. (b) The dependence of the linewidth and the peak ionisation rate on $\alpha$. (c) Persistence time of the power broadening. A \SI{375}{\uW} off-resonance laser pulse is first applied in each measurement cycle. After a time delay of $t_\textrm{w}$, a \SI{9}{\uW} resonant pulse is applied for spectral scan. (d) Peak ionisation rates from CW (red circle) and pulsed (brown diamond) experiments on the effective optical power. The solid line shows a linear fit. Inset: a zoom-in plot of the bottom left region.}
	\label{fig:fig3}
\end{figure}

First, the spectrum is measured under different resonant powers while the total power remains fixed. Figure~\hyperref[fig:fig3]{3(a)} shows three example spectra measured with the same total optical power of \SI{41}{\micro\watt} but different $\alpha$ values, where the linewidth has been broadened to about \SI{85}{\MHz}. All the spectra can be well fitted with a Lorentzian lineshape. The fitted peak ionisation rate and linewidth are plotted as a function of $\alpha$, where $\alpha$ represents the proportion of resonant optical power to the total optical power, in Fig.~\hyperref[fig:fig3]{3(b)}. The peak ionisation rate increases linearly with the resonant optical power. The linewidth shows no obvious trend and fluctuates around \SI{85}{\mega\hertz}, which is consistent with the linewidth shown in Fig.~\hyperref[fig:fig2]{2(d)}. Therefore, we conclude that the power broadening solely depends on the total optical power regardless of the resonant proportion.

Second, the persistence time of the power broadening is studied with a two-pulse experiment. A \SI{375}{\uW} off-resonance laser pulse is first applied, and a linewidth of over \SI{150}{\MHz} is expected according to the results in Fig.~\hyperref[fig:fig2]{2(d)}. After a time delay of $t_\textrm{w}$, the Er spectrum is scanned with a resonant pulse to examine any persisting power broadening due to the first laser pulse. The power of the resonant pulse is \SI{9}{\uW}, which is low enough to avoid inducing any additional broadening, and the pulse length of \SI{4}{\us} ensures a sufficient ionisation count rate and a reasonable measurement time for each spectrum. Figure.~\hyperref[fig:fig3]{3(c)} shows the spectra with time delays ranging from \SI{50}{\ns} to \SI{10}{\us}. All the spectra show a uniform linewidth of \SI{34(4)}{\MHz}, consistent with the linewidth observed in Fig.~\hyperref[fig:fig2]{2(d)} under the same power. This observation indicates that the persistence time of the power broadening is shorter than a few microseconds.

Overall, the power broadening increases significantly with the total optical power used for Er spectral scan but does not depend on the resonant proportion of fixed total power, and the persistence time of the power broadening is shorter than a few microseconds. Consequently, the power broadening observed in this device is not due to instantaneous spectral diffusion. It could however be caused by other spectral diffusion processes on a fast time scale~\cite{bottger_optical_2006}, such as light-induced charging and discharging of the traps in the device~\cite{Photoluminescence_1996,Robinson2000,Abbarchi2008}. When the device is illuminated under a higher optical power, stronger charge fluctuations can lead to a stronger broadening of the Er\tplus transition via the Stark effect. As a result, the laser light at the peak centre frequency cannot drive the Er\tplus transition effectively when the instantaneous transition frequency diffuses away from the centre frequency. As a verification, the peak ionisation rate observed at different powers is plotted against the effective power in Fig.~\hyperref[fig:fig3]{3(d)}. The effective power is defined as $P_\mathrm{eff} = \alpha \times P \times W_\mathrm{min} / W(P)$, where $\alpha = 0.496$, $P$ is the total optical power, $W_\mathrm{min}$ is the observed minimum linewidth of \SI{32}{\MHz}, and $W(P)$ is the observed linewidth at the optical power of $P$. As shown in Fig.~\hyperref[fig:fig3]{3(d)}, peak ionisation rates from both CW and pulsed measurements follow a linear fit (solid line) well, consistent with power broadening due to spectral diffusion.

In this work, the spectral scanning speed is limited by the ionisation count rate to the time scale of 10 minutes. The spectral diffusion on this slow time scale appears smaller than \SI{15}{\mega\hertz}. Further investigation on spectral diffusion on a broader range of time scales can be conducted on devices that show faster ionisation rates. Correlation measurements~\cite{sallen_subnanosecond_2010} can also be used to achieve a finer time resolution.

\section{Conclusion}
In summary, we present an effective method of using the Er-induced photoionisation process to investigate spectroscopic properties of a single Er\tplus ion in Si. The ionisation rate is found to be an appropriate quantity to represent the optical transition probability for spectroscopic studies, and the single ion spectra display a Lorentzian lineshape at all optical powers in use. The linewidth of the single Er\tplus ion remains near constant at optical powers below \SI{5}{\micro\watt} and increases monotonically as the optical power increases up to \SI{61}{\micro\watt}. In addition, the broadening does not change significantly with the proportion of resonant power under a fixed total optical power and shows a persistence time below a few microseconds. The results indicate power broadening induced by spectral diffusion on a fast time scale supported by the linear dependence of the peak ionisation rate on the effective optical power.

This method provide a more consistent spectral lineshape than the asymmetric lineshape used in the high-precision strain measurement~\cite{ZhangQ2019} and would enhance the strain measurement precision. However, the power broadening creates challenges for coherent experiments such as optical Rabi oscillations. Also, the observed minimum linewidth of \SI{32(2)}{\MHz} in this device is considerably larger than the sub-megahertz homogeneous linewidths that have been measured for Er\tplus ions in bulk Si~\cite{berkman_sub-megahertz_2021}. These could be limited by charge fluctuations in the device's channel or interface, via the Stark effect~\cite{bartholomew_optical_2017} or magnetic dipole-dipole interactions. Therefore, effect of device size and structural variations, controlled doping~\cite{PRXQuantum21}, and charge depletion~\cite{ChristopherElectrical} on the broadening mechanisms needs to be further investigated so that approaches to obtaining reduced linewidths can be devised.
\begin{acknowledgments}
This work was supported by the National Key R\&D Program of China (Grant No. 2018YFA0306600), Anhui Initiative in Quantum Information Technologies (Grant No. AHY050000), and Natural Science Research Foundation of Anhui Province (Grant No. 2108085MA15).
\end{acknowledgments}

\bibliography{Er_APS}
\end{document}


\preprint{APS/123-QED}

\title{Supplemental Material for Spectral broadening of a single Er$\ensuremath{^{3+}}\xspace$ ion in a Si nano-transistor}
\maketitle
\date{\today}

\section{Decay time of \texorpdfstring{E\MakeLowercase{r}}{Er} excited state}
\begin{figure}[h]
	\centering
	
	\includegraphics[width=.7\linewidth]{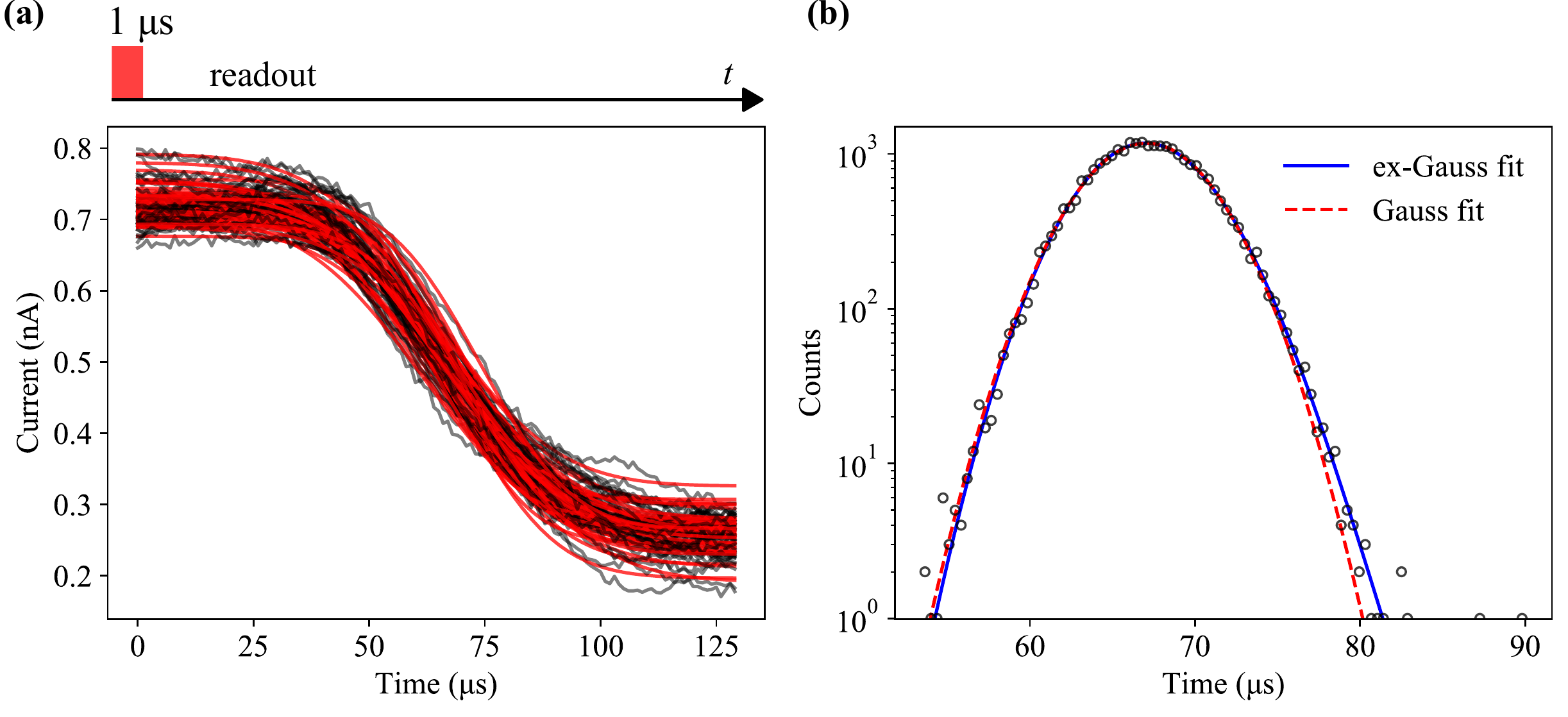}
	\caption{\label{fig:figS1}Decay time of the \ere excited state. (a) Pulse sequence and example current-time traces from successful ionisation cycles. (b) Histogram of all current switching times, fitted to an exponential-Gaussian function (solid line) and a  Gaussian function (dashed line).}
\end{figure}
Due to the electric bandwidth limitation (10~\si{\kilo\hertz}), the decay time of the \ere excited state can not be measured accurately in this work. However the upper limit of the decay time can be determined from the variation of the current switching time after a short excitation pulse~\cite{hu_time-resolved_2022}. Fig.~\hyperref[fig:figS1]{S1(a)} shows a number of current-time traces after a \SI{1}{\us} resonant pulse. Each current trace is fitted to a step function, and a switching time, the start time of the falling edge, is extracted. A histogram of all switching times is plotted in Fig.~\hyperref[fig:figS1]{S1(b)} and is fitted to a Gaussian function and an exponential-Gaussian function. The histogram well follows a Gaussian distribution, consistent with the previous observation~\cite{hu_time-resolved_2022}. The exponential-Gaussian fitting gives a decay time of 0.72$\pm$\SI{0.10}{\us}. Additionally, the \SI{1}{\us} pulse length also sets an upper limit on the detectable decay time. Therefore, these results indicate an upper limit of  \SI{1}{\us} on the decay time of the Er \ere state.
\section{Monte-Carlo Simulation}
A three-state model is used to simulate the excitation and relaxation processes of the Er\tplus ion as well as the ionisation and reset of the single trap. The three states are \{\erg, 0\}, \{\ere, 0\}, and \{\erg, 1\}, where \erg or \ere stands for the electronic state of the Er\tplus ion, and 0 or 1 stands for the neutral state or the ionised state of the single trap, respectively.
The transitions between these three states are modelled using a Monte-Carlo method with multiple sets of parameters. For each set of parameters, the occurrence times of ionisation events and reset events under continuous-wave (CW) laser excitation are recorded. These events are equivalent to the detected events in a current--time trace. Similar to the data analysis in the main manuscript, an ionisation rate, $\nu_\textrm{i}$, and a reset rate, $\nu_\textrm{r}$ can be calculated from statistics of these events.
In the simulations, the range of excitation rate, $\gamma_\textrm{e}$, is from 1~kHz  to 1$\times$10$^3$~kHz . Different $\gamma_\textrm{i}\slash\gamma_\textrm{ni}$ ratios are chosen to show their impact on the ratio of $\nu_\textrm{i}$ to $\gamma_\textrm{e}$. The reset rate does not affect the ionisation rate and can be experimentally measured. Therefore, the reset rate is chosen at  \SI{12.5}{\kHz} for all the simulations.

\begin{figure}[h]
	\centering
	\includegraphics[width=0.7\linewidth]{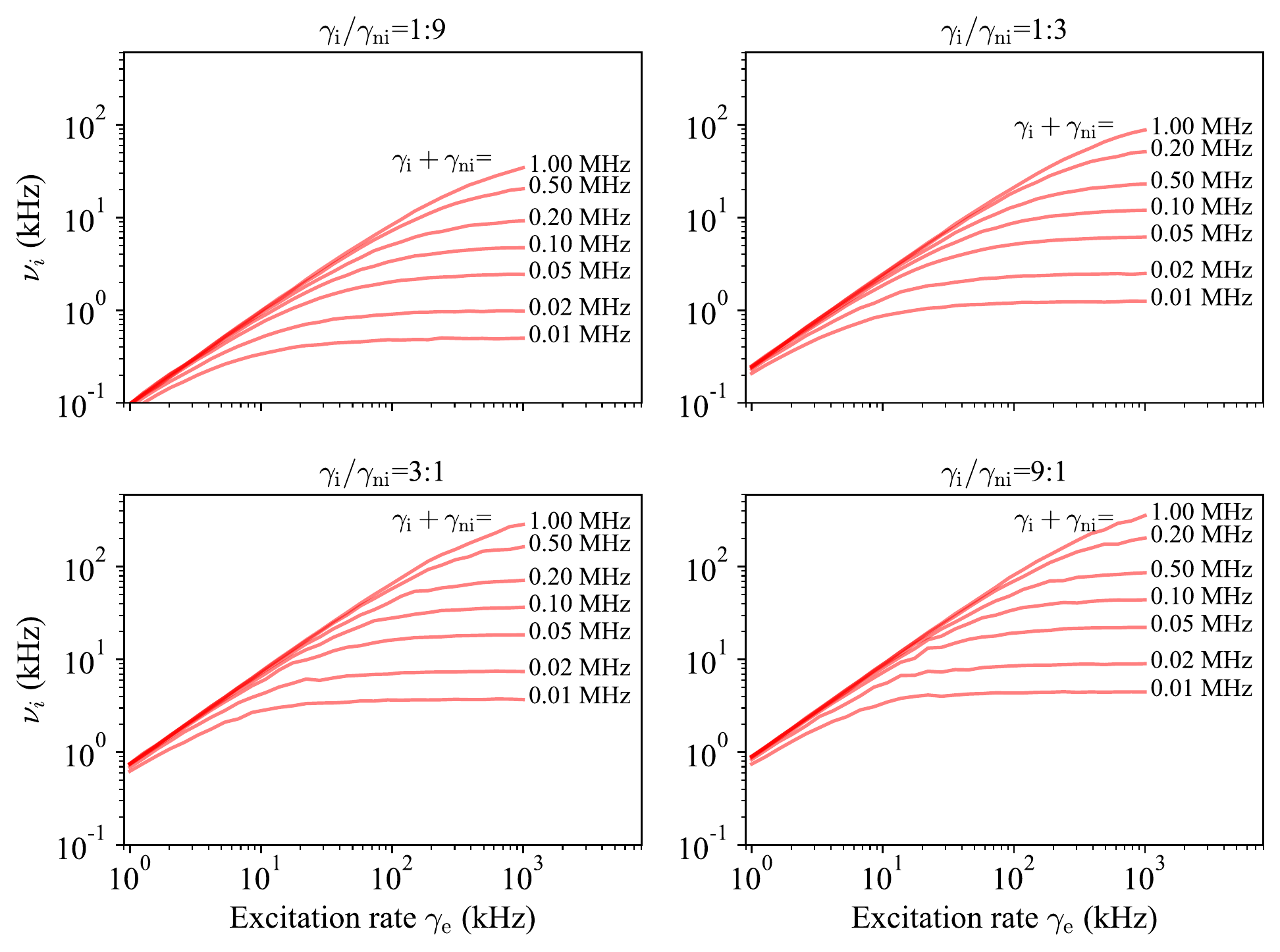}
	\caption{\label{fig:figS2}Dependence of ionisation rate, $\nu_\textrm{i}$, on the excitation rate, $\gamma_\textrm{e}$, from Monte-Carlo simulations.}
\end{figure}

The simulation results are shown in Fig.~\hyperref[fig:figS2]{S2}. For all these parameter combinations, $\nu_\textrm{i}$ increases linearly with $\gamma_\textrm{e}$ until $\gamma_\textrm{e}$ approaches the total decay rate, $\gamma_\textrm{i}+\gamma_\textrm{ni}$. A higher $\gamma_\textrm{i}\slash\gamma_\textrm{ni}$ ratio contributes to a higher ionisation decay proportion, and thus leads to a higher ratio of $\nu_\textrm{i}$ to $\gamma_\textrm{e}$ as shown in Fig.~\hyperref[fig:figS2]{S2}.
\section{Two-pulse measurement of the reset rate}
 
\begin{figure}[h]
	\centering
	\includegraphics[width=0.5\linewidth]{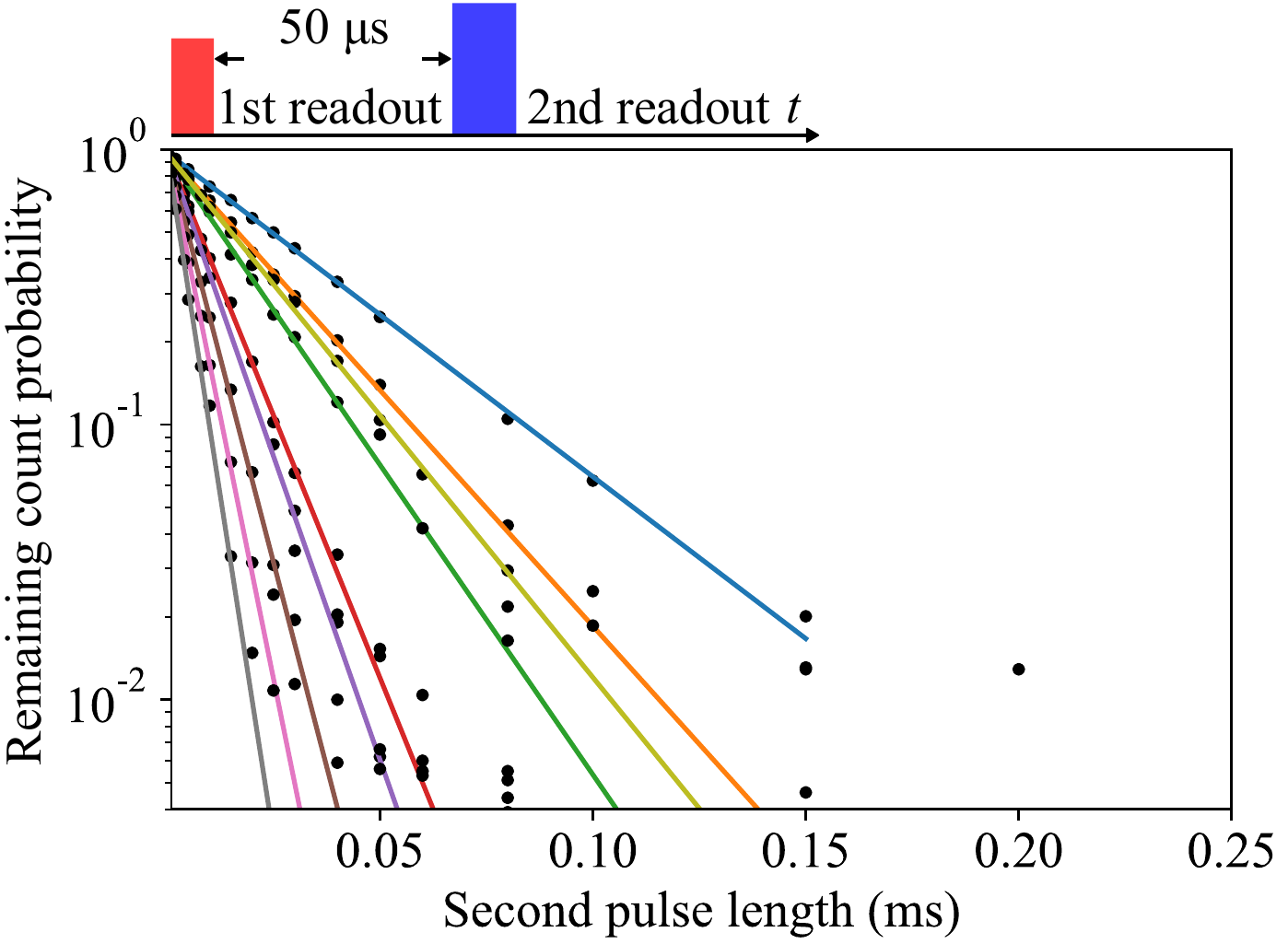}
	\caption{\label{fig:figS3} Light-assisted reset rates. Pulse sequence (top panel) and persisting ionisation probability after the second pulse in the two-pulse experiments under high optical powers.}
\end{figure}

The pulse sequence is shown on the top section of Fig.~\hyperref[fig:figS3]{S3}. A resonant pulse is first applied in each cycle to generate photoionisation events. The current signal in the `1st readout' window is compared against a threshold level to determine whether there is an ionisation count. At the end of the `1st readout' window, an off-resonance pulse is applied to reset any ionisation signal. Whenever an ionisation count appears in the `1st readout' window of a cycle, the current signal in the beginning of the `2nd readout' window is compared against the threshold level to determine if the ionisation count has been reset by the end of the second laser pulse. The ratio of the remaining counts in the beginning of the `2nd readout' window to the counts that appear in the `1st readout' window is defined as the remaining count probability. This measurement is carried out with different lengths and optical powers of the second pulse. As shown in Fig.~\hyperref[fig:figS3]{S3}, the remaining count probability drops exponentially as the second-pulse length increases. Each coloured line represents an exponential decay fit of the results under one second-pulse power. The fitted reset rates are shown in Fig.~2(b) in main manuscript.
 
\newpage
\section{Spectra with absolute ionisation rates}
\begin{figure*}[h]  
	\centering
	\includegraphics[width=1\linewidth]{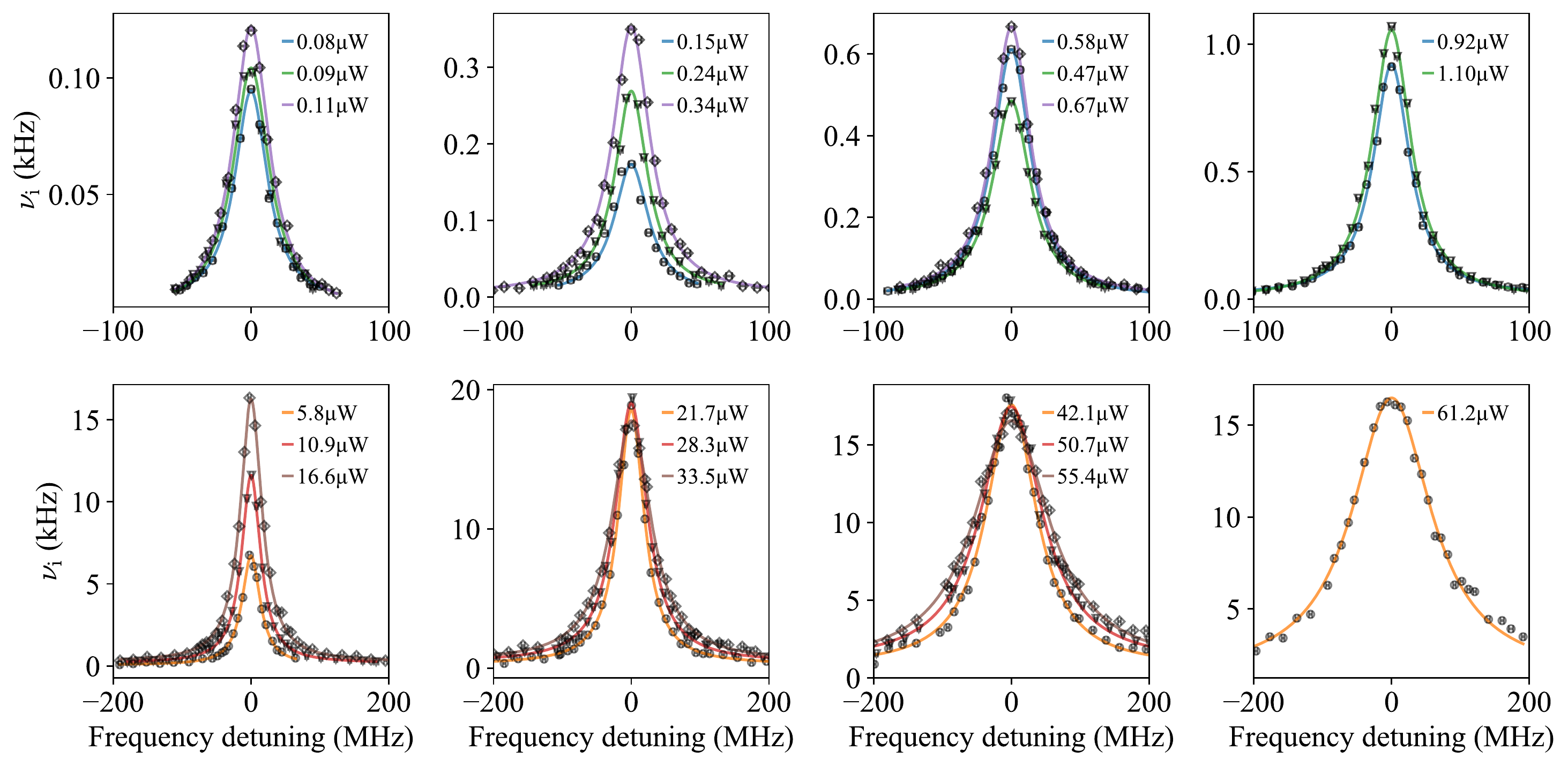}
	\caption{\label{fig:figS4}Spectra replotted with absolute ionisation rates, $\nu_\textrm{i}$. The first row corresponds to the CW spectra shown in Fig.~1(g) in the main manuscript, and the second row corresponds to the pulsed spectra in Fig.~2(c).}
\end{figure*} 
\bibliography{Fr_APS_supp}